\documentclass[doublecol,figures]{epl2} 

\usepackage{amssymb}

\title{Capillary condensation in an active bath}
\shorttitle{Title} 

\author{M. Kne\v zevi\'c \and H. Stark}
\shortauthor{M. Kne\v zevi\'c and H. Stark}

\institute{                    
  Institut f\" ur Theoretische Physik, Technische Universit\" at Berlin - Hardenbergstra\ss e 36,
  D-10623 Berlin, Germany\\
}


\abstract{
We study capillary condensation in a bath consisting of active Brownian particles
(ABPs) and the forces acting on the capillary close to the motility-induced phase separation (MIPS). The capillary is modelled as two parallel rods of high aspect ratio which are fixed in space. We consider a bath of ABPs having a self-propulsion speed much larger than the critical speed necessary for MIPS to occur. For a given particle speed, we gradually increase the packing fraction of ABPs, starting from a homogeneous dilute phase of ABPs and going towards the binodal of MIPS. Depending on the packing fraction of ABPs and capillary width, we find that the effective force between the capillary rods can be either attractive or repulsive. In fact, with increasing width it shows damped oscillations as long as capillary condensation occurs. We analyze them in detail by studying the distribution of particle distances from the inner and outer wall of the capillary, respectively. In addition, we examine the capillary in the active bath prepared under conditions close to the critical point. We do not observe signs of the presence of long-range Casimir interactions.       
}

\begin{document}

\maketitle

\section{Introduction}

Unlike passive systems, which consist of particles displaying directed motion only in the presence
of external fields, active systems\cite{cates12,viscek12,marchetti13,aronson13,bechinger16,zoettl16} encompass a collection of self-driven units. They have the ability to convert energy typically stored in the environment into autonomous dissipative and directed motion. Due to their steady energy consumption, active systems are fundamentally out of equilibrium. Examples of active matter systems include an assemblage of living creatures covering different length scales, ranging from the macroscopic scale, where sheep herds\cite{ginelli15}, bird flocks\cite{ballerini08}, or fish schools\cite{katz11} occur, to microscopic scale with collections of motile bacteria or sperm cells\cite{lauga09,ishikawa09,elgeti15,yeomans14}. Galvanised by the alluring physics of living active systems, scientists have produced and examined the collective dynamics of numerous experimental realisations of artificial active systems, comprising, but not limited to, autophoretic colloids\cite{theurkauff12,buttinoni13,palacci10,anderson89,kapral13,juelicher09,golestanian05,golestanian07,golestanian09,howse07,pohl14,stark18}, shaken grains\cite{deseigne10}, colloidal surfers\cite{palacci13} and rollers\cite{bricard13}. Notably, spherical active colloids\cite{fily12,redner13,zheng13,bialke12,tenhagen11,vanteeffelen08,fily14,yang14,stenhammar14}, which possess the simplest architecture of these artificial systems, are conveniently accessible for theoretical modelling.

\begin{figure}
	\onefigure[scale=0.45]{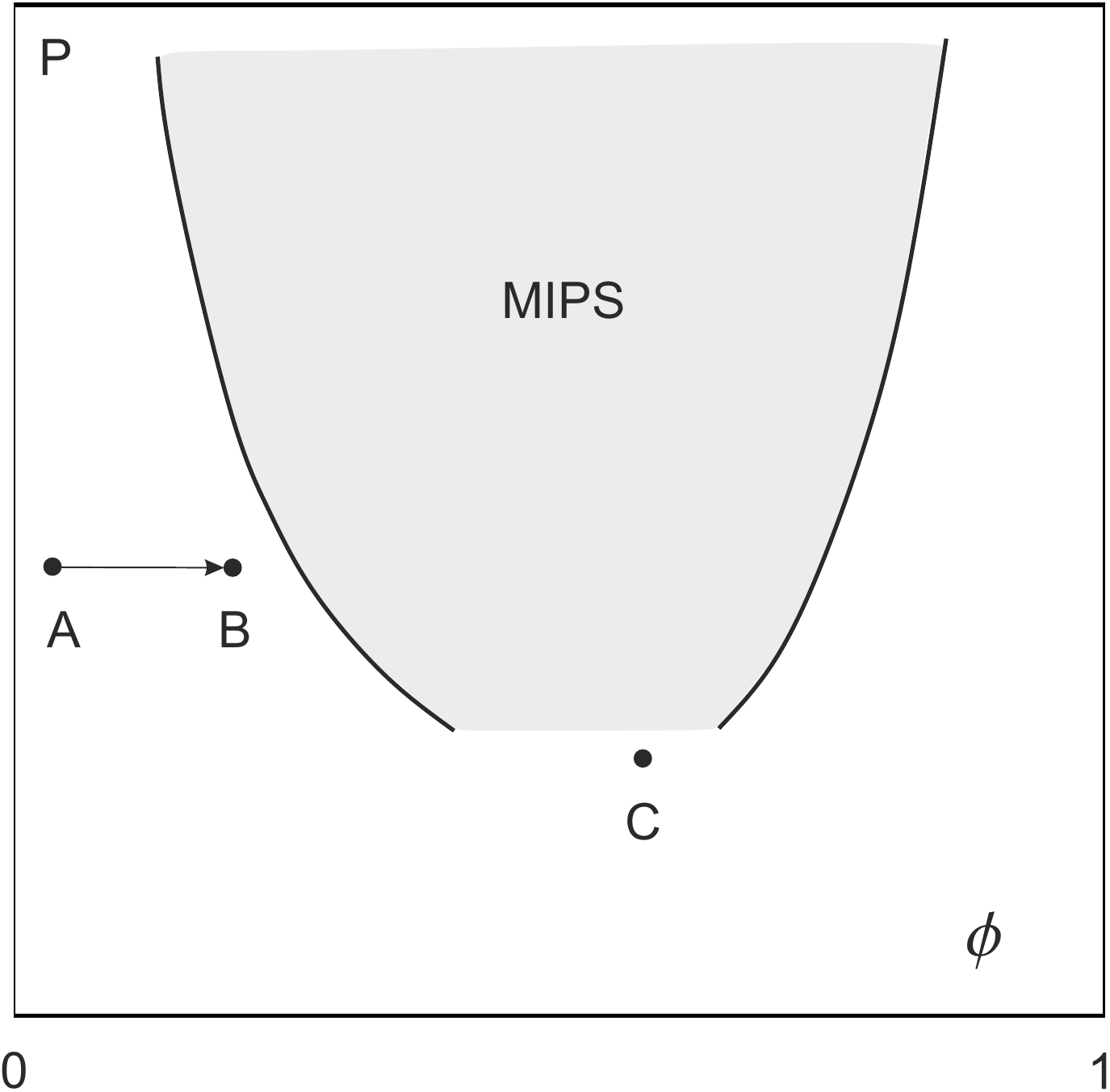}
	\caption{A sketch of the phase diagram for the system of interacting ABPs; here $\phi$ is the bulk packing fraction of active particles, and $P \sim v$ is their persistence number. Solid lines denote binodals of MIPS (shaded region). Points A and B are located in the region of homogeneous dilute phase and have equal persistence number $P$; C denotes 
		the critical point\cite{siebert18}.}
	\label{diagram}
\end{figure}

In this article, we concentrate on interacting active Brownian particles (ABPs)\cite{schweitzer03,romanczuk12} in 2D, which is a well established model for self-propelled spherical colloidal particles immersed in a solvent. The pairwise interactions between the colloids are purely repulsive and isotropic. We consider only steric interactions and neglect hydrodynamic effects\cite{kim13,zoettl14,blaschke16}. In contrast to dilute systems of passive colloids with repulsive interactions, ABPs can phase-separate\cite{fily12,redner13,speck14,bialke15,bialke15-2,speck16} into dilute and dense domains for a sufficiently large self-propulsion speed. This phenomenon has been termed motility-induced phase separation (MIPS)\cite{tailleur08,tailleur09,cates13,cates15,stenhammar13} and corresponds to the shaded region of the phase diagram shown in fig.~\ref{diagram}. Outside of this region the system is found in a homogeneous state, either dilute or dense. MIPS originates from a dynamic instability\cite{bialke13,speck16} which emerges when the mean time between particle collisions is sufficiently smaller than the decorrelation time of directed motion.

Capillary condensation\cite{rouquerol99} is a paramount phenomenon in both naturally appearing and synthetic porous arrangements. It develops when multilayer adsorption of vapour molecules into a porous medium takes place, eventually proceeding to the state characterised by pores substantially filled with condensed liquid. Initially, vapour molecules are adsorbed by the pore surface. The wetting layers grow in size and ultimately drive the capillary bridging and the formation of liquid menisci. The exceptional property of capillary condensation is the occurrence of vapour condensation below the saturation vapour pressure of the bulk liquid.
\begin{figure}
	\onefigure[scale=0.30]{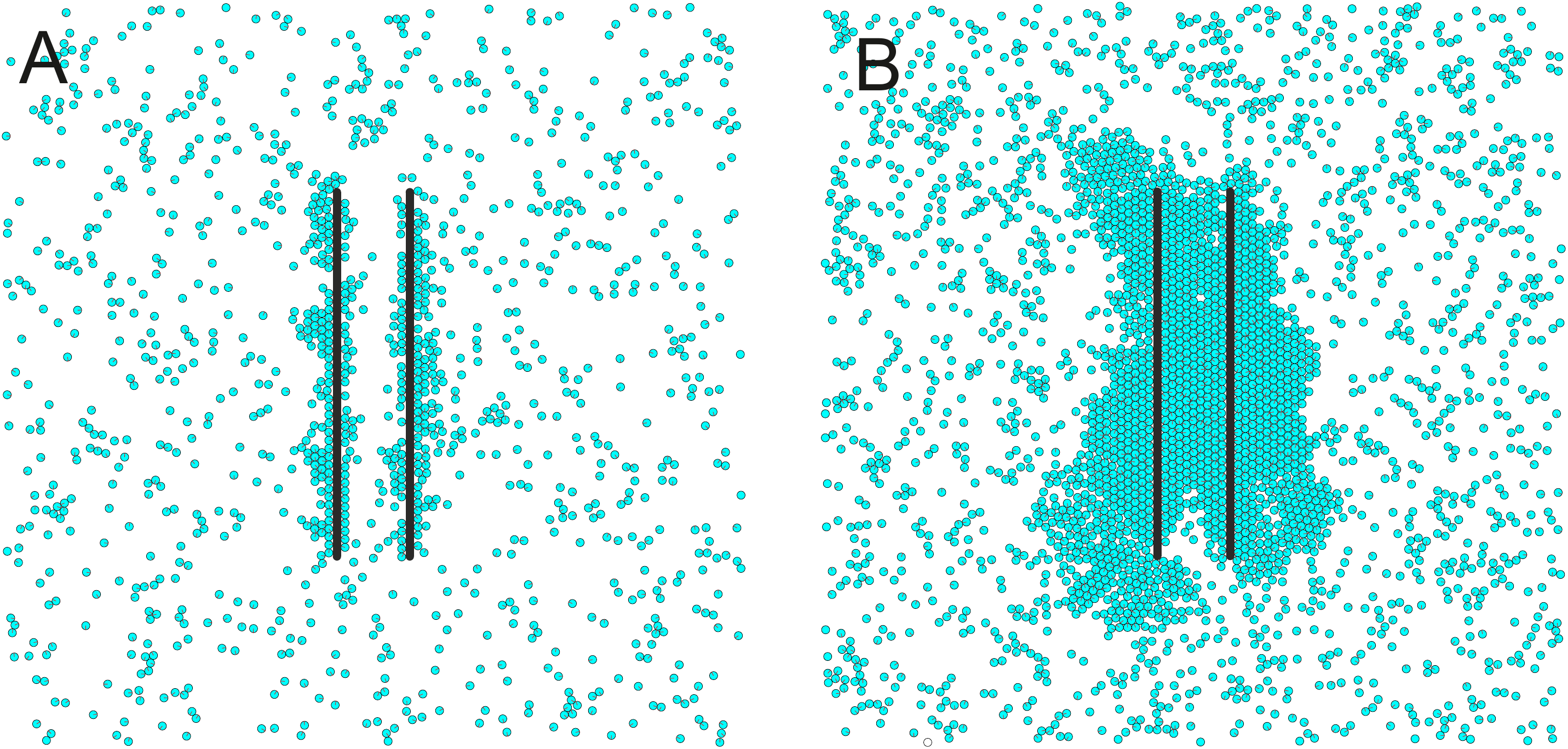}
	\caption{Capillaries in an active bath. Typical simulation snapshots for capillary width $w/\sigma = 10$: (A) $\phi \approx 0.08$, (B) $\phi \approx 0.16$. Here we only show a part of the simulation box in the vicinity of the capillary. The other parameters are specified in the main text.}
	\label{snapshots}
\end{figure}

Motivated by similarities between the liquid--gas phase separation and the dense--dilute phase coexistence of active particles, we study capillary condensation in a bath consisting of ABPs. Previous works have already examined interactions between passive bodies immersed in an active bath, either of rodlike\cite{ni15,ray14}, spherical\cite{harder14} or general asymmetric\cite{baek18} shape. However, they did not explore what happens close to the binodal of MIPS. Thus, in analogy with the ordinary vapour--liquid capillary condensation, we examine a capillary immersed in a bath prepared in a homogeneous dilute phase. The capillary consists of two long parallel rods pinned in space at a fixed distance between them. When an active particle collides with a surface, it needs some typical reorientation time to point away from the surface and escape. Thus, in contrast to passive particles, active particles pile up at walls even in the absence of attractive particle--wall interactions\cite{li09,volpe11,enculescu11,elgeti13,schaar15}. For a given particle speed, we increase the packing fraction of ABPs, following the path A--B in fig.~\ref{diagram} towards the binodal line. Active particles tend to accumulate at capillary walls and form wetting layers of dense phase; the closer one gets to the binodal, the thicker the wetting layers. For high enough bulk packing fractions, one obtains a capillary filled with ABPs. We measure the effective force between the capillary rods as a function of the distance between them and the packing fraction of the ABPs. It turns out that this force oscillates between attraction and repulsion with increasing distance\cite{ni15}. The range of the oscillations extends to larger rod separations when approaching the binodal. We study the behaviour of these forces in detail and, in particular, discuss the underlying physical mechanism governing the change of force sign.

\section{Model}

We consider a system of $N$ interacting active Brownian particles (ABPs) in two dimensions, which self-propel with a constant speed $v$ and have a mobility $\mu$. The dynamics of ABPs is described by overdamped Langevin equations
\begin{eqnarray}
	\label{poseq}
	\dot{\mathbf{r}}_i &=& v \mathbf{u}_i - \mu [\sum_{j \neq i} \nabla_{\mathbf{r}_i}  V(\mathbf{r}_{i} - \mathbf{r}_j)
	+ \nabla_{\mathbf{r}_i}  V_\mathrm{C}], \\
	\label{orteq}
	\dot{\theta}_i &=& \sqrt{2D_\mathrm{R}}\xi_i.
\end{eqnarray}

Here $\mathbf{u}_i \equiv (\cos \theta_i, \sin \theta_i)$ is the unit orientation vector of particle $i$, $D_\mathrm{R}$ denotes its rotational diffusivity, and $\xi_i$ is Gaussian white noise of zero mean and unit variance, $\langle \xi_i(t) \xi_j(t') \rangle = \delta_{ij} \delta (t-t')$. We perform numerical simulations in the regime of large propulsion speed which allows us to neglect the effect of translational diffusivity in the above equations. Active particles interact with each other through pairwise forces, which are given by the negative gradient of the Weeks-Chandler-Andersen (WCA) potential
$$
\label{potential}
V(\mathbf{r}) = \left\{
\begin{array}{ll}
4 \varepsilon \left [ \left (\frac{\sigma}{|\mathbf{r}|} \right )^{12} - \left (\frac{\sigma}{|\mathbf{r}|} \right )^{6}\right ] + \varepsilon, & \quad |\mathbf{r}| \leq 2^{1/6} \sigma \\
0, & \quad |\mathbf{r}| > 2^{1/6} \sigma
\end{array}
\right.
$$
Here $\epsilon$ is the strength of the potential and $\sigma$ its characteristic length (where the potential takes the value $\varepsilon$).
We carry out simulations in a 2D box of size $L_x \times L_y$ and use periodic boundary conditions. 

In the bath of interacting active particles we immerse a capillary and pin it to a fixed position in space. The capillary is modelled as two long rods of length $h$, positioned at a distance $w$ between them, which we refer to as the capillary width. The rods are placed parallel to the $\hat{\mathbf{y}}$ axis, and their centres are located at positions $\mathbf{r}_{1\mathrm{c}}=(L_x/2-w/2,L_y/2)$ and $\mathbf{r}_{2\mathrm{c}}=(L_x/2+w/2,L_y/2)$.
The interactions of active particles with the capillary are accounted through the potential $V_\mathrm{C} = V_1(\mathbf{r}_i - \mathbf{r}_{1i}) + V_2(\mathbf{r}_i - \mathbf{r}_{2i})$, where
$V_1$ and $V_2$ are the contributions from the two rods, respectively.
They are both of the WCA form, meaning that the rods have an effective thickness equal to the effective diameter of an active particle. Here, $\mathbf{r}_{1i}$ and $\mathbf{r}_{2i}$ are the points located along the axis of the rods that are closest to the active particle $i$.

We use $\sigma$ as the unit of length, persistence time $\tau_\mathrm{R} = D_\mathrm{R}^{-1}$ of an active particle as the unit of time, and we measure energies in units of $k_\mathrm{B}T$, where $T$ is the temperature of the solvent surrounding active particles. We take $D_\mathrm{T}= \sigma^2 D_\mathrm{R}/3$, where $D_T$ is the translational diffusion constant. This relation holds in 3D, but we assume its approximate validity in a 2D system as well. We introduce the persistence number $P=l_\mathrm{p}/\sigma$. It quantifies the persistence length $l_\mathrm{p}=v\tau_\mathrm{R}$, which is the distance an active particle travels in roughly the same direction.
The eqs.~(\ref{poseq}) and (\ref{orteq}) can then be cast into a dimensionless form, with two independent parameters: the persistence number $P$ and the potential strength $\varepsilon/k_\mathrm{B}T$. The area fraction of active particles, $\phi = \frac{N\sigma^2\pi}{4L_xL_y}$, together with the persistence number $P$ determine the phase diagram of active particles in the bulk; for a schematic see fig.~\ref{diagram}. Active particles separate into dense and dilute regions for large enough packing fractions and self-propulsion speeds.
\begin{figure}
	\onefigure[scale=0.29]{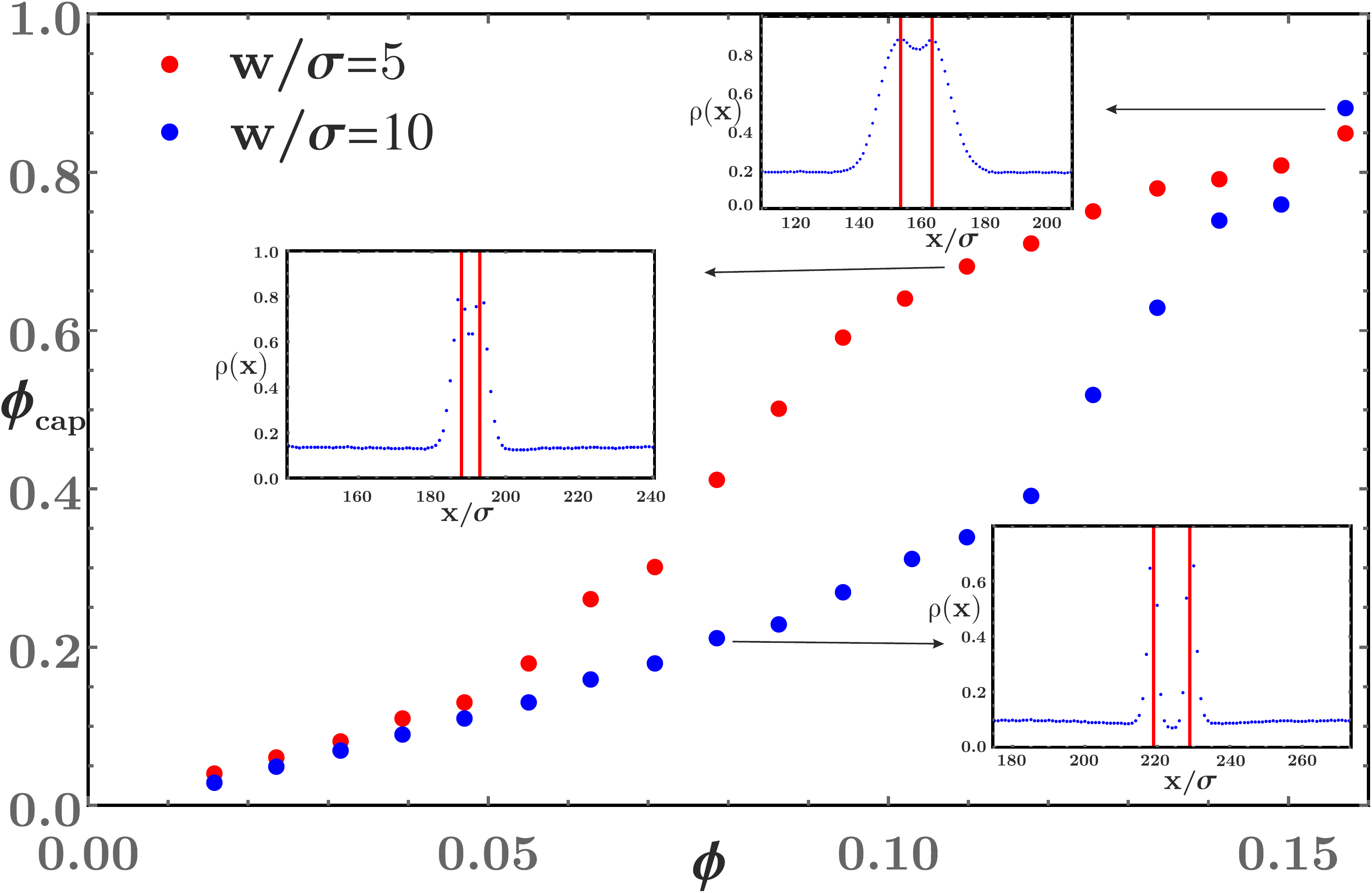}
	\caption{Packing fraction of active particles inside the capillary $\phi_\mathrm{cap}$ as
		a function of the bulk packing fraction $\phi$ for two capillary widths $w$. Insets: density of active particles along $x$ for three selected points of the main figure; the bin size is $\Delta x/\sigma =1$ and the data are averaged in the layer $y \in [L_y/2-h/6, L_y/2+h/6]$. The positions of capillary rods are marked with red vertical lines.}
	\label{dens_phi}
\end{figure}

In the following, we study capillary condensation in the active bath. Starting from low packing fractions $\phi$, corresponding to a homogeneous dilute phase of active particles (point A in the phase diagram of fig.~\ref{diagram}), we increase the value of $\phi$ towards the point B lying in the vicinity of the binodal of MIPS. At some point along the line A--B active wetting of capillary walls starts to occur. The wetting layers of dense phase of ABPs grow in size until the capillary is eventually filled up with active particles. For a fixed persistence number $P$, we measure the total forces $F_1$ and $F_2$ on the capillary rods, exerted by active particles along the direction $\hat{\mathbf{x}}$ perpendicular to the rods:
\begin{eqnarray}
\label{force1}
F_1 &=& \sum_i \nabla_{x_i} V_1(\mathbf{r}_i - \mathbf{r}_{1i}), \\
\label{force2}
F_2 &=& \sum_i \nabla_{x_i} V_2(\mathbf{r}_i - \mathbf{r}_{2i}).
\end{eqnarray}

We work in a slab geometry with $L_x = 2 L_y$. We fix the number of particles to $N=10000$ and the simulation box size $L_x \times L_y$ is adjusted to obtain the required packing fractions $\phi$. We use $\varepsilon/k_\mathrm{B}T = 100$ and $h/\sigma = 50$. The time step is $\delta t = 10^{-5} \tau_\mathrm{R}$, and simulations are typically run up to times $t = 3000 \tau_\mathrm{R}$.

\section{Capillary condensation force -- changing packing fraction $\phi$}

We fix the persistence number of active particles to $P= 26.67$, which is twice as large as
the estimated critical value $P_\mathrm{c}$ for MIPS. For this value of $P$ the binodal point was found to be at $\phi \lessapprox 0.2$; see for example the phase diagram in fig.~2(d) of ref. \cite{bialke15-2}, where the P\'eclet number $P_\mathrm{e} = v \sigma/D_\mathrm{T}\equiv 3 P$ is used as control parameter instead of $P$. 

Due to symmetry, the forces $F_1$ and $F_2$ acting on the two rods are on average of equal magnitude and opposite sign: the case $F_1<0$, $F_2>0$ corresponds to repulsion, and $F_1>0$, $F_2<0$ to attraction between the rods. In the following we thus report our results only for the force $F_2$.

For two capillary widths $w/\sigma = 5$ and $10$, we perform a number of simulations with different bulk packing fractions $\phi$ of active particles. We measure the time averaged packing fraction of particles inside the capillary $\phi_{\mathrm{cap}} = \frac{n \sigma^2 \pi}{4hw}$ and a dimensionless capillary pressure $p_2 = \frac{F_2 \mu \sigma}{vh}$; here $n$ is the number of particles inside the capillary region of area $h \times w$ and $p_2$ is the net force $F_2$ exerted on the second rod rescaled by the force $vh/(\mu \sigma) $ to stop a dense layer of active particles pushing against one side of the rod. The reduced number density $\rho = \frac{N\sigma^2}{L_x L_y}$ is varied within the range $0.02 \le \rho \le 0.2$, leading to packing fractions $\phi = \rho \pi/4$ which encompass very dilute active baths on one end and active baths close to their binodal point of phase separation on the other. For $w/\sigma = 10$ typical simulation snapshots are shown for $\phi \approx 0.08$ and $0.16$ in fig.~\ref{snapshots}; see also corresponding movies A and B in the SI. In the case $\phi \approx 0.16$ the capillary is mostly filled up with active particles, although occasional partial emptying of the capillary can temporarily occur (see movie B). 

Fig.~\ref{dens_phi} shows the packing fraction $\phi_\mathrm{cap}$ inside the capillary as
a function of $\phi$ for two selected capillary widths.
For low $\phi$ both capillaries contain only a small population of active particles inside them.
As $\phi$ is increased, active wetting of capillary walls gradually starts to occur, leading to thick wetting layers for high enough $\phi$. The steeper increase of $\phi_\mathrm{cap}$ with $\phi$
then indicates the onset of capillary condensation. For a wide range of $\phi$, the capillary of smaller width $w/\sigma = 5$ naturally reaches higher values of $\phi_\mathrm{cap}$ compared to the capillary with $w/\sigma = 10$ since capillary condensation already occurs at smaller densities. Nevertheless, at $\phi \approx 0.16$ both capillaries are fully packed with particles. Typical density profiles of active particles are shown as insets in fig.~\ref{dens_phi}.

\begin{figure}
	\onefigure[scale=0.27]{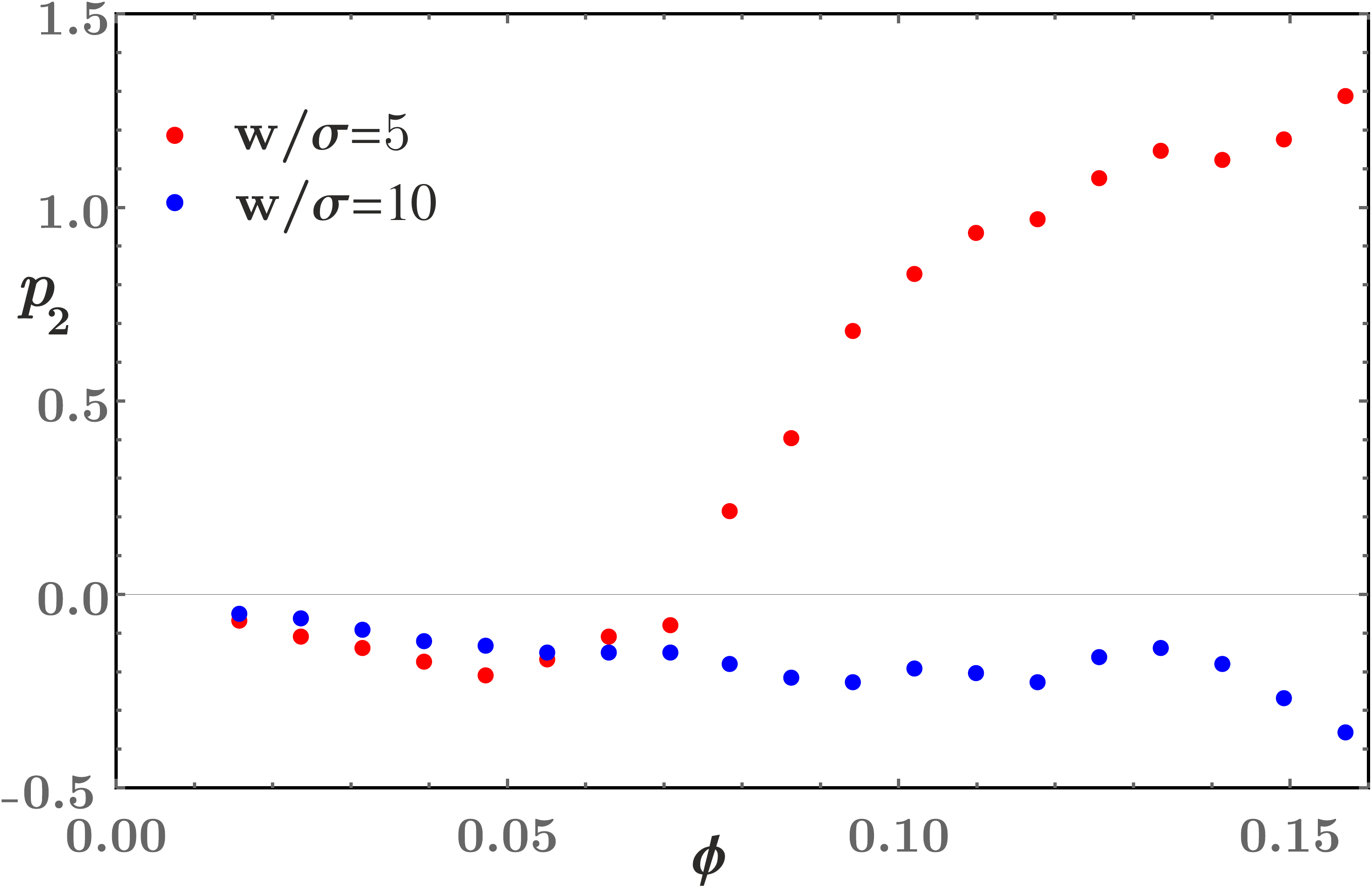}
	\caption{Dimensionless capillary pressure $p_2$ as a function of the bulk packing fraction $\phi$ for two capillary widths $w$.}
	\label{p_phi}
\end{figure}

Unlike $\phi_\mathrm{cap}$, which displays growth with $\phi$ for both widths $w$, the capillary pressure $p_2$ shows a remarkably distinct behaviour for different $w$, see fig.~\ref{p_phi}. For $w/\sigma = 5$, and very low $\phi$ there is an effective attraction between the capillary rods ($p_2<0$). This is rationalised in the following way. Each rod has two walls, the one facing the inside of the capillary, which we term the inner wall, and the other facing the bulk of active-particle bath, which we term the outer wall. For very dilute baths, the outer wall has a greater chance to be hit by an active particle, compared to the inner wall, whose exposure to the active bath is screened\cite{ni15,ray14} by the presence of the other rod. As the net force on the rod is given by a difference of forces acting on its inner and outer walls, this screening clearly leads to an effective attraction between the rods. The attractive force increases with $\phi$ up to $\phi \approx 0.05$. At this point, the behaviour of $p_2(\phi)$ changes and the magnitude of attraction weakens. At $\phi \approx 0.08$ the capillary on average has $\phi_\mathrm{cap} = 0.4$, the interactions between active particles and inner walls become more frequent, leading to a crossover to an effective repulsion between the rods ($p_2>0$). This repulsion further increases in magnitude as we go towards the binodal of MIPS, and consequently to a capillary fully packed with active particles.

In stark contrast to the previous case, the wider capillary, $w/\sigma = 10$, exhibits an attractive force between the rods for all $\phi$. Thus, at $\phi \approx 0.16$, both capillaries are densely packed with active particles, but the narrower one exhibits strong repulsion, while the wider one displays attraction. As the outer walls of both capillaries are exposed to the same active bath (characterised by $P$ and $\phi$), they exhibit the same average force independent of the capillary width. The change in force for different $w$ then stems solely from the change in force on the inner capillary walls. This motivates us to measure the capillary force for a wide range of capillary widths, which is the topic of the next section.

\section{Capillary condensation force -- changing capillary width $w$}
\begin{figure}
	\onefigure[scale=0.29]{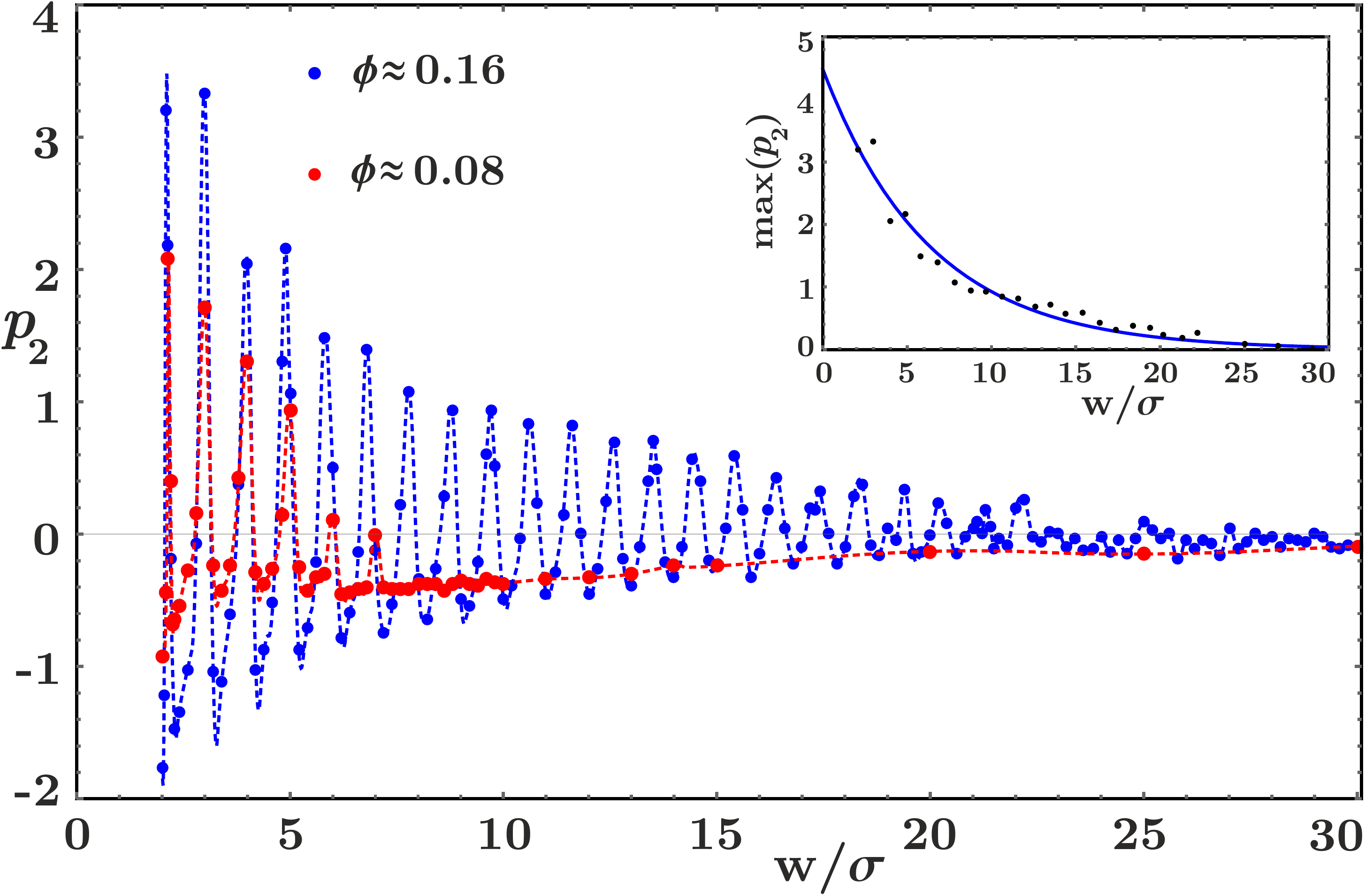}
	\caption{Dimensionless capillary pressure $p_2$ as a function of capillary width $w$ for two selected bulk packing fractions $\phi$ (blue and red points). The dashed lines are a guide to the eye. Inset: The maxima of $p_2$ for $\phi \approx 0.16$ are shown in black. The blue line is a fit to the exponential form $ae^{-w/\xi}$, where $\xi = \xi(\phi)$.}
	\label{p_w}
\end{figure}

We select two bulk packing fractions $\phi \approx 0.08$ and $0.16$, and vary the capillary width
in small increments within the range $2\le w/\sigma \le 30$. The dimensionless capillary pressure $p_2$ as a function of $w$ is shown in fig.~\ref{p_w}. At lower packing fraction $\phi \approx 0.08$, the pressure $p_2$ alternates between negative and positive values\cite{ni15} as the width $w$ is increased, starting from $w/\sigma = 2$ up to $w/\sigma \approx 7$. Active particles cannot enter a very narrow capillary of width $w/\sigma = 2$, meaning that there are only interactions between particles and outer capillary walls, which leads to a pure attractive force between capillary rods ($p_2<0$). As the distance between the rods is slightly increased to $w/\sigma = 2.15$, the particles have enough space to fit inside the capillary. Yet, under these conditions, the intrusion of particles inside the capillary interior occurs only seldomly. Those particles that manage to get inside are typically stuck in there for a long time and frequently interact with inner capillary walls, leading to an increase in the force on the inner wall. As a consequence, the pressure $p_2$ switches sign from negative to positive, and also exhibits a jump in its magnitude by a factor of about $2$. With further increase of width, $w/\sigma = 2.25$, the particles inside the capillary have more space to wiggle around, which leads to a weakening of interactions with the inner walls. Thus the pressure drops and the rods again exhibit an effective attraction. The oscillations of pressure persist up to $w/\sigma \approx 7$ and have a period of roughly $w = \sigma$. The oscillations are asymmetric with respect to the $p_2$--axis: the maxima of $p_2$ are much larger than the absolute values of corresponding minima. The magnitude of maxima and minima decreases with $w$ and at $w/\sigma \approx 7$ there is a crossover to a new regime: the pressure $p_2$ becomes negative for all $w/\sigma >7$. This change coincides with a drop of the packing fraction of particles inside the capillary $\phi_\mathrm{cap}$: for $2<w/\sigma <7$ the packing fraction is on average $\phi_\mathrm{cap} \approx 0.5$, while for $w/\sigma >7$ it drops to $\phi_\mathrm{cap} \approx 0.3$ and has a tendency to further decrease with increasing $w$ (plot not shown). The drop of $\phi_\mathrm{cap}$ leads to a decrease in the number of encounters of active particles with the inner walls of the capillary and consequently accounts for the effective attraction between the capillary rods. Finally, for very large capillary widths $w/\sigma \gg 30$ the packing fraction $\phi_\mathrm{cap}$ decays to the bulk value $\phi$ and the effective force between the rods vanishes, as both the inner and outer walls of rods are exposed to the bulk of active bath.

At higher packing fraction $\phi \approx 0.16$, the pressure oscillations are now visible up to 
$w/\sigma \approx 22$, see fig.~\ref{p_w}. The longer range of oscillations is due to the high packing fraction inside the capillary: for $2<w/\sigma<22$ one has on average $\phi_\mathrm{cap} \approx 0.8$. For widths $w/\sigma>22$, where $p_2$ mostly assumes small negative values, the packing fraction $\phi_\mathrm{cap}$ decays slowly towards the bulk packing fraction $\phi$ with increasing $w$, leading to $p_2 \rightarrow 0$. The periodicity of oscillations, $w=\sigma$, seems to be the same as in the case of $\phi \approx 0.08$. At $\phi \approx 0.16$, collisions of active particles with inner and outer capillary walls are more frequent, which leads to larger magnitudes of maxima and minima of $p_2$ compared to the corresponding magnitudes at $\phi \approx 0.08$. According to our findings, the maxima of $p_2$ display a simple exponential decay with $w$, see the inset of fig.~\ref{p_w}. The absolute values of the minima of $p_2$ also follow a similar form.

\begin{figure}
	\onefigure[scale=0.29]{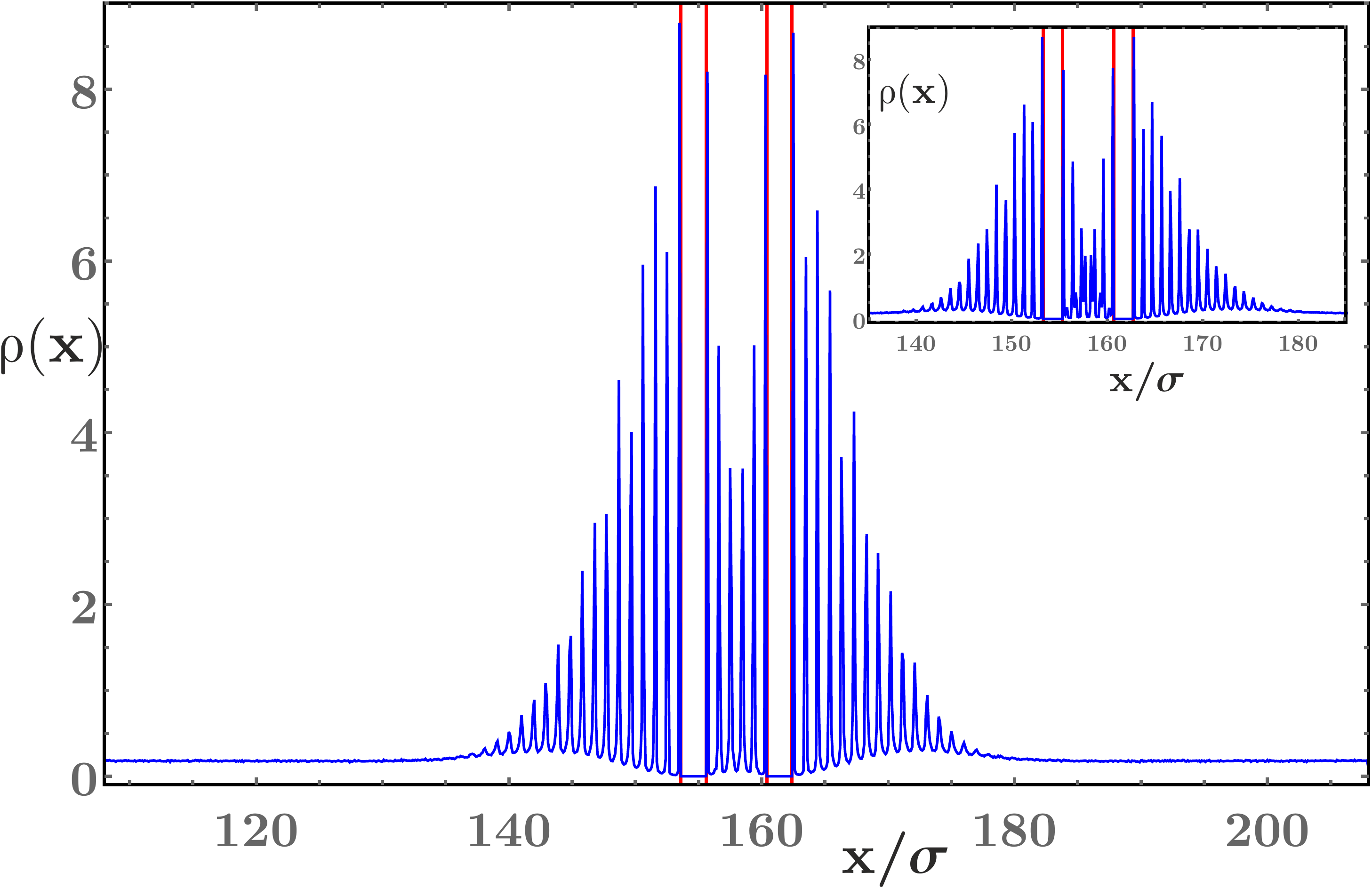}
	\caption{Density of active particles along $x$ for $\phi \approx 0.16$ and $w/\sigma = 6.8$; the bin size is $\Delta x/\sigma = 0.1$ and the data are averaged in the layer $y \in [L_y/2-h/6, L_y/2+h/6]$. The positions of capillary rods are marked with red vertical lines. Inset: density of active particles along $x$ for $w/\sigma = 7.4$; all other parameters are the same.}
	\label{layers}
\end{figure}

The change of sign of $p_2$ stems from subtle packing effect of active particles inside the capillary due to the formation of particle layers. To examine this effect in more detail, we focus on the case $\phi \approx 0.16$. We choose two capillary widths $w/\sigma = 6.8$ and $w/\sigma = 7.2$, which correspond to a maximum of $p_2$ and one of its nearest neighbour minima, respectively. The density $\rho(x)$ of active particles inside the capillary and in the immediate vicinity outside the capillary shows layering (see fig.~\ref{layers}). It is interesting that $\rho(x)$ inside the capillary displays approximately the same quantitative behaviour in both cases, $w/\sigma = 6.8$ (see fig.~\ref{layers}) and $w/\sigma = 7.2$ (figure not shown). Only with further increase of $w$ starting from the minimum of $p_2$ at $w/\sigma = 7.2$, one observes an increment in the number of layers inside the capillary.
As the inset of fig.~\ref{layers} demonstrates, for $w/\sigma = 7.4$ additional layers/peaks form, which then evolve into a new 7th layer in the center of the capiilary, when reaching the next maximum of the repulsion force. Note again that the layering is very dynamic since partial emptying of the capillary temporarily occurs (see movie B).
\begin{figure}
	\onefigure[scale=0.44]{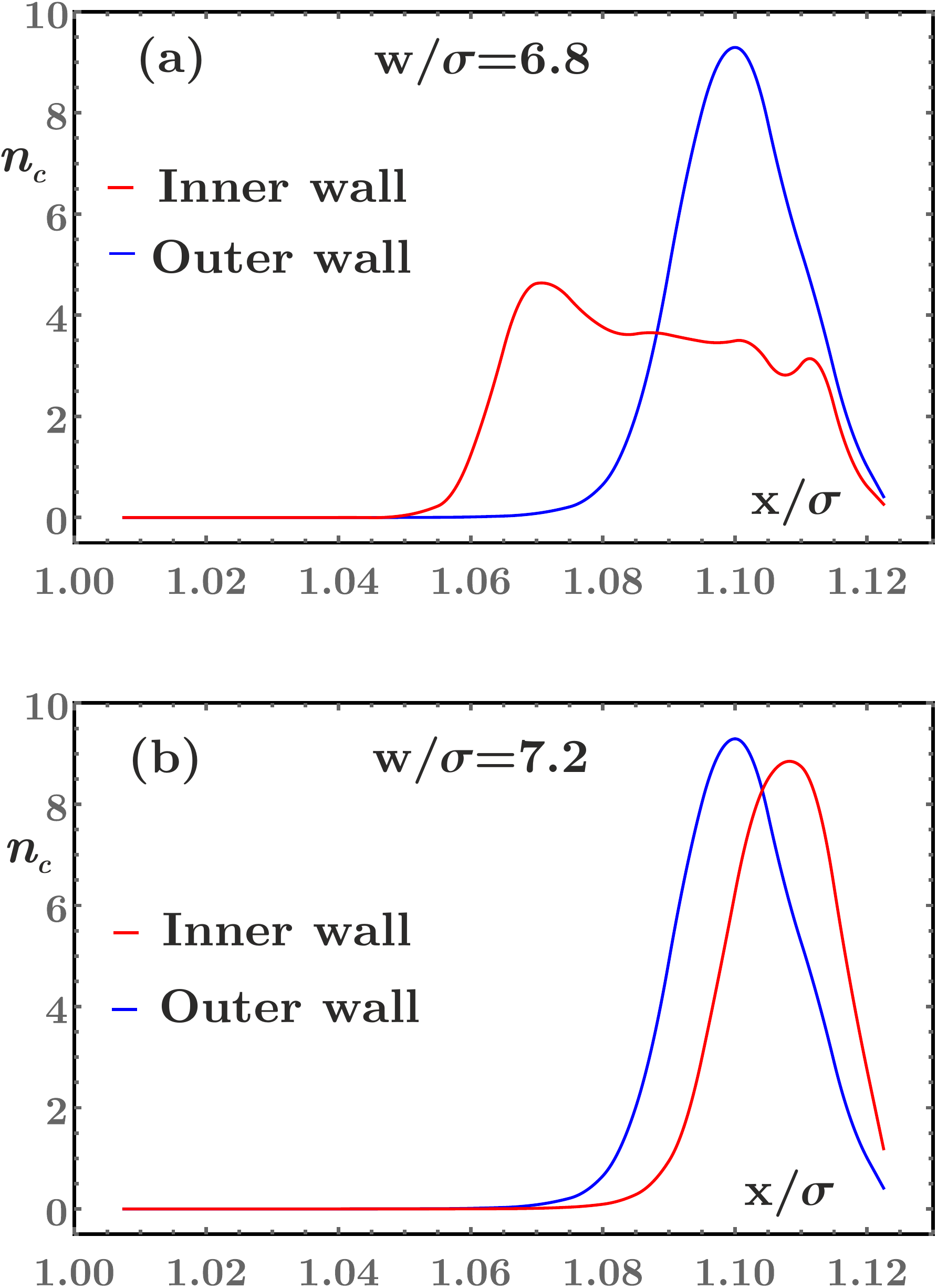}
	\caption{Average number of particles $n_\mathrm{c}$ found at a distance $x$ from the inner (red curve) and outer (blue curve) capillary wall: (a) $w/\sigma = 6.8$, (b) $w/\sigma = 7.2$; in both cases $\phi \approx 0.16$.}
	\label{nc68}
\end{figure}

Given that there is no significant difference in layering of active particles for capillaries of widths $w/\sigma = 6.8$ and $w/\sigma = 7.2$, and that the force between active particles and rods is short-ranged, the change of sign of $p_2$ should be connected to the distribution of particle distances from the inner capillary wall.
Thus, we take particles that interact with capillary rods and compute the time-averaged distributions of particle--rod distances for both the inner and outer wall of a rod. They are shown in fig.~\ref{nc68} for $w/\sigma = 6.8$ and $w/\sigma = 7.2$, respectively. The particles and rods interact with each other if the distance between their centres is smaller than the cut-off distance $2^{1/6}\sigma \approx 1.122 \sigma$. As expected, the distribution of particle--outer wall interactions does not depend on the capillary width $w$ (blue curves in fig.~\ref{nc68}). The distribution is peaked around $x/\sigma \approx 1.10$. Note that for our choice of parameters the force to stop an active particle $v/\mu$ is equal to $80\frac{k_\mathrm{B}T}{\sigma}$. The WCA force matches this value at a particle--rod distance $x_\mathrm{s}/\sigma = 1.11$. The small discrepancy between $x_\mathrm{s}$ and the position of the distribution peak is probably due to particles in the additional layers pushing against the particles in the first layer directly in contact with the outer wall. At $w/\sigma = 7.2$ the distribution of particle--inner wall distances is shown in red in fig.~\ref{nc68}(b). It is peaked at $x/\sigma \approx 1.11$. Note that for all $x/\sigma < 1.105$ the distribution of particle--outer wall distances takes higher values compared to the distribution of particle--inner wall distances. The corresponding particles contribute more to the total force on the
rod, thus leading to an effective attraction between the rods, i.e. $p_2<0$ for $w/\sigma = 7.2$. 

In fig.~\ref{nc68}(a), for $w/\sigma = 6.8$ the corresponding distribution for the particle--inner wall distances consists of multiple peaks which are merged together, and the highest of them is found at $x/\sigma \approx 1.07$. For all $x/\sigma < 1.088$ the distribution of particle--inner wall distances now assumes larger values compared to the other distribution. Thus, at width $w/\sigma = 6.8$ we obtain an effective repulsion between capillary rods, i.e. $p_2>0$. 
In conclusion, the distance between capillary rods together with the surface enforced layering and the softness of the WCA potential determine how strongly active particles interact with the inner capillary walls and thereby the sign of the effective force between the rods.

\section{Capillary in a critical active bath}

\begin{figure}
	\onefigure[scale=0.29]{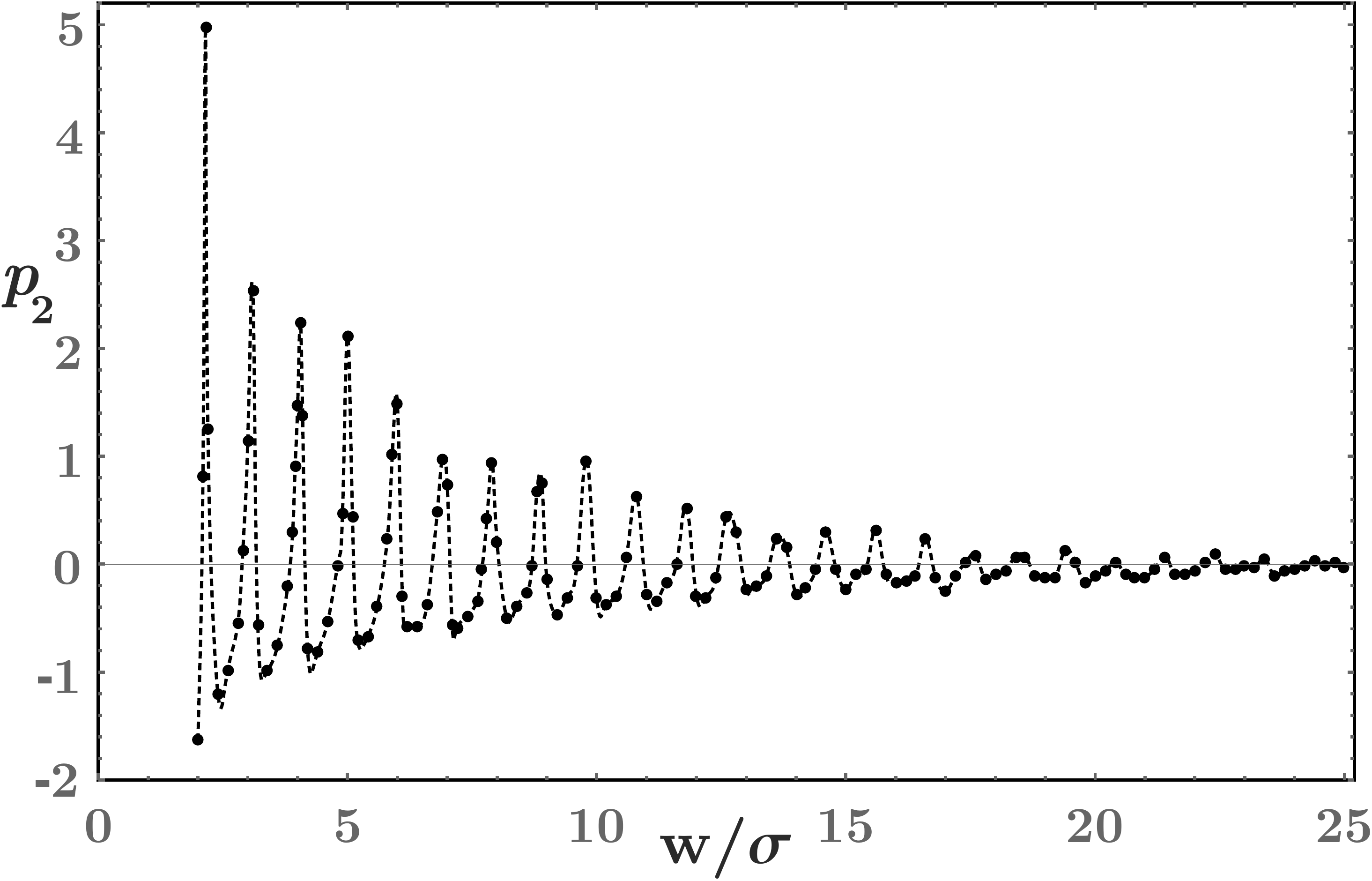}
	\caption{Dimensionless capillary pressure $p_2$ as a function of capillary width $w$ 
	in a bath of ABPs prepared in the vicinity of the critical point of MIPS (black points); 	
	parameters: $\phi \approx 0.5$ and $P = 13.33$ . The dashed line is a guide to the eye.}
	\label{cpoint}
\end{figure}

We now prepare the active bath in the vicinity of the critical point of MIPS, see point C in fig.~\ref{diagram}. The location of the critical point was previously estimated\cite{bialke15-2} to be at $P_\mathrm{cr} = 13.33$ and $\phi_\mathrm{cr} \approx 0.6$ for an infinite active bath. Here we use a lower value $\phi_\mathrm{cr} \approx 0.5$ to adjust for the finite size of our simulation box. For a capillary immersed in a critical active bath one might expect the emergence of Casimir long-range interactions\cite{casimir48,fisher78,gambassi09} between the capillary rods. According to our results presented in fig.~\ref{cpoint}  the pressure $p_2$ displays the same qualitative behaviour as in the case $P=26.67$ and $\phi \approx 0.16$. The oscillations still have a periodicity of approximately $w = \sigma$ and a range of about $25 \sigma$. The maxima of $p_2$ are larger compared to the ones of fig.~\ref{p_w}. This difference can be attributed to the fact that the critical bath is characterized by a higher bulk packing fraction $\phi_\mathrm{cr} \approx 0.5$. Thus, we do not see a clear sign of Casimir interactions.

\section{Summary}
We studied capillary condensation in an active bath. Using numerical simulations we analysed the effective forces between the capillary walls in a number of different active bath settings, including very dilute baths, baths close to the binodal line of MIPS, and the bath being in the state close to the critical point. In all cases the effective force on the rods constituting the capillary oscillates between positive and negative values as long as capillary condensation occurs, otherwise the force is attractive. While the periodicity of the oscillations is determined by the layering of the active particles in the capillary, both the range of force and the magnitude of its maxima and minima depend on the active bath properties, namely the persistence number $P$ and the bulk packing fraction $\phi$ of active particles.

Our research sheds light on the subtle interactions of inclusions immersed in an active bath, which are strongly determined by the activity of its constituents. It will help exploring active baths for assembly processes on the microscopic scale.

\acknowledgments
The authors thank Thomas Speck for helpful discussions. MK gratefully acknowledges support from the Alexander von Humboldt Foundation through a postdoctoral research fellowship.





\end{document}